The Thermodynamics of the living organisms: entropy production in the cell 1

# THE THERMODYNAMICS OF THE LIVING ORGANISMS: ENTROPY PRODUCTION IN THE CELL

By: Araceli Venegas Gómez

## 0. Abstract

Trying to identify the entropy production within a cell has been part of debates and studies in the last century. First the idea was to make a resemblance of a cell with a Carnot engine, which is the most thermodynamically perfect machine. This approach was clearly not the best, since the yield achieved within a cell cannot be ideal, but can we even measure it? Several models approach the living cell, since the very simple one (e.g. Prigogine model) to more elaborated proposals. The concept of entropy has been the centre of discussions within several scientific fields. To interpret how entropy is produced in the complicated system of a cell is as hard as to understand how life originated at the first place.

Understanding the way a cell works is key in biology, medicine, and multiple other scientific fields. Thermodynamics is essential in multitude of processes around us. Trying to identify the entropy production within a cell has been past of debates and studies in the last century. I give here an insight of what has been done and a personal opinion about the topic, focusing in the results from Himeoka and Kaneko paper.

## 1. Thermodynamics of life: introduction

Thermodynamics studies the energy flow, heat and movement, in structures among the universe. When the study concerns a living system, i.e. a cell, it is usual to refer to open system thermodynamics or nonequilibrium thermodynamics. That is mainly because the concepts of flow, growth and change are not static, not in equilibrium.

When looking to the presence of life it is normal to think about order in nature, and not of the universal tendency for things to fall into chaos and disorder. To measure this "disorder" we need to understand what "entropy" means [1].

## 2. General concept of Entropy

Entropy is a concept that even though it is well described physically and mathematically there is still kind of a mysterious halo about it. It is well known that entropy is related to the measurement of disorder, and in our universe in all irreversible process the entropy is always positive [1-3].

Entropy can be defined macroscopically as well as microscopically, being the last one by means of the statistical mechanics.

Macroscopically the entropy (S) for a thermodynamically reversible process is defined by the following equation as the heat (Q) into the system divided by the uniform temperature (T).

$$S = \frac{Q}{T} \qquad (1)$$

But knowing that entropy is a function of state and it can be considered as by means of reversible transformations through a trajectory the change in entropy is defined as:

$$\Delta S = \int \frac{\delta Q}{T} \qquad (2)$$

In the modern microscopic interpretation of entropy in statistical mechanics, entropy is a logarithmic measure of the number of states ($W_i$) with significant probability of being occupied:

$S = k_B ln W_i$ ;

being k the Boltzmann constant = 1,38x10$^{-23}$ J K$^{-1}$

## 3. Entropy in living cells

It has been since the 1940s that the concept of entropy was differentiated when applied to life from other forms of matter organization. Living organisms are highly organizational and therefore it seems that it feeds from "negative entropy" or, by other words, maintaining and getting to a stationary condition where the entropy level is low.

Nevertheless it is necessary to understand that the proper definition of the second law of thermodynamics says that the entropy of an adiabatically isolated system never decreases. In this context a living cell or organism is not an isolated system, since it gets the nutrients from the exterior, that is, there is an exchange of heat or matter with the environment, and by doing so, we have to consider the system as an open one, together with its environment, restoring the balance to the universe in an increase of entropy, or by other words, increase of disorder [1-4].

## 4. Thermodynamic potentials

In order to represent the thermodynamic state of a system we use scalar quantities called thermodynamic potentials. Hereafter there is a table that summarizes all these potentials:

| Name | Symbol | Formula | Definition/concept |
|---|---|---|---|
| Internal Energy | U | $\int \left( TdS - pdV + \sum_i \mu_i dN_i \right)$ | Energy contained within the system |
| Helmholtz free energy | F | U - TS | Energy available in the system to realize useful work |
| Enthalpy | H | U + pV | Internal energy of the system plus the product of pressure and volume of the system |
| Gibbs free Energy | G | U + pV -TS | Is the maximum amount of non-expansion work that can be extracted from a closed system |

Table 1: Thermodynamic Potentials. Source: [3]

## 5. The Carnot cycle

The Carnot cycle is a theoretical thermodynamic cycle proposed in the XIX century and it is also called the ideal machine, since it is the most efficient cycle for converting a given amount of thermal energy into work (or creating a temperature difference by doing a given amount of work). Let´s imagine that we have an ideal gas inside a cylinder with a piston so that the system can exchange energy with two thermal focuses $T_1$ and $T_2$. In the figure 2 the four transformations of the cycle are described [1-3].

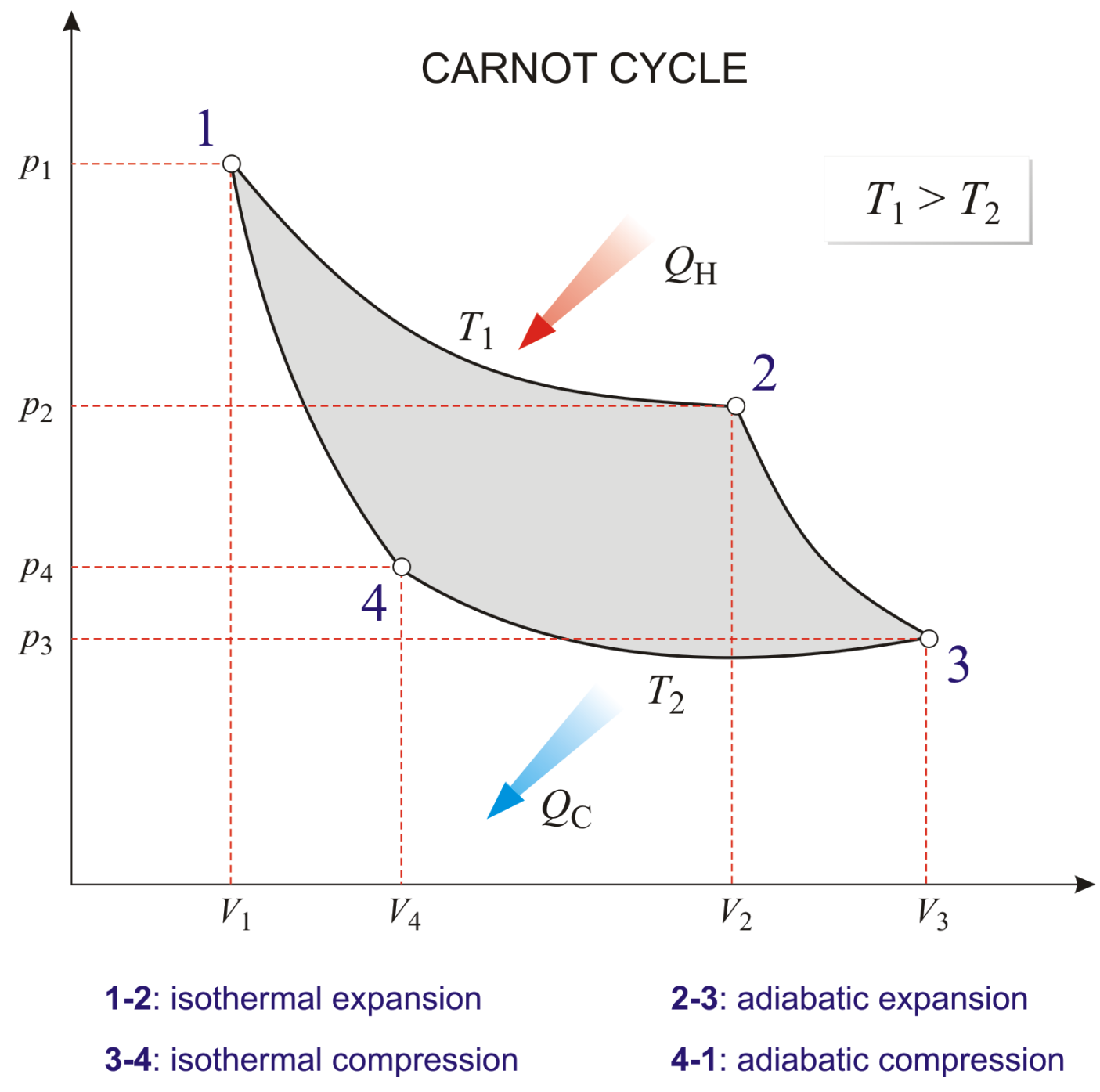

Figure 2: The Carnot Cycle. Source: [www.periodni.com]

The main question now is to understand the main differences between a cell and the standard Carnot engine. Following the paper from Himeoka and Kaneko [2] there are three main differences: “cells contains catalysts, second, the catalysts are synthetized within the cell as a result of catalytic reactions, and third, cell volume growth results from membrane synthesis from nutrient components, facilitated by the catalyst”. Let´s try to analyse these three points, first of all assuming the analogy of a cell to the Carnot cycle, as indicated by Smith [5], “with the substitution of particle number for heat, and chemical potential for temperature in the Carnot diagram”.

To summarize these points, we can understand the cell as an open system, exchanging chemicals and energy through the environment, which is indeed obtained from the nutrients through catalytic reactions. Therefore I would like to summarize it, as “a cell is an open system exchanging energy through catalytic reactions”, and this is the main difference with the ideal Carnot engine, which is in total equilibrium with the environment. Nevertheless I would like to support the argument of Davies [6]: “the catalysts needed for its operation must be generated internally”, which doesn’t contradict my point.

The minimal entropy production in a living cell is not found at a quasi-static level limit as in the standard Carnot engine but at a non-zero nutrient uptake rate. In order to understand these results it is necessary to explain how the thermodynamics works in biological and chemical processes and our level of understanding so far through the use of simple cell models.

# 6. Introduction to the Non linear Thermodynamics of biological and chemical processes

Mostly of the systems that we find in the nature are changing over time, and continuously subject to flux of matter and energy through chemical reactions. That is why instead of approaching entropy in living systems as a measure of disorder we need to take entropy as the flow of energy in a system.

## 6.1 Irreversible Thermodynamics

Irreversible thermodynamics is based on the Gibbs formula and the evaluation of the entropy production and flow.

The first assumption for the irreversible processes in nature states the principle of locality, meaning, that all interactions are only between nearby atoms or neighbouring molecules.

Secondly, to consider the flow of energy it is necessary to understand the concept of flow. When there is a flow it means there is a force, which comes from the difference (gradient) of a variable.

The speed of entropic production can be described as:

$$S = JF \qquad (3)$$

where $J = \phi F$ ; with J the flow, F the force and $\phi$ the transport coefficient.

Then taking the relation between both, $S = \phi F^2$ → as the entropy production must be not negative, the transport coefficient must be positive.

In the case of coupled flows, the condition that the production entropy must be positive doesn't entail that each of the members implicated are positive. These kind of entropic compensations are very common in biological systems [3,7].

### 6.1.1 Onsager Theorem

The Onsager relations express the equality of certain ratios between forces and flows in non-equilibrium thermodynamics but with the notion of local equilibrium. When having different flows due to different forces ($F_1$, $F_2$, … $F_n$) we can consider [3]:

$$J_i = \sum_{j=1}^{n} \phi_{ij} F_j \qquad (4)$$

with $\phi$ the transport coefficient matrix:

$$\phi_{ij} = \lim_{F_j \to 0} \left(\frac{\partial J_i}{\partial F_i}\right)_{F_{i \neq j}=0} \qquad (5)$$

which, according to Onsager, is symmetric: $\phi_{ij} = \phi_{ji} \; \forall \; i \neq j$

#### 6.1.2 Entropy production equation

Let´s consider the following expression for the entropy balance equation:

$$\frac{\partial s}{\partial t} + \nabla J_s = \sigma \qquad (6)$$

being $\sigma$ the entropy production [7], which has to be always ≥0.

Now, the entropy production can be decompose in two parts, one related to the change of forces (x) and the other one to the change of flows (j):

$$d\sigma = d_x\sigma + d_j\sigma = \sum_k J_k dX_k + \sum_k X_k dJ_k \qquad (7)$$

In the whole domain of the validity of the thermodynamics for irreversible processes the most general result whenever the boundary conditions are time-independent $d_x\sigma \leq 0$ .

### 6.2 Non-linear thermodynamics

In the case of chemical reactions linear approximations are not enough, since some coefficients may vary. This variation must be considered non-linear [8].

The relation $\frac{\partial s}{\partial t} \leq 0$ can be extended in order to take into account flow processes in inhomogeneous systems over the volume of the system. In order to include these forces and flows (e.g force to compensate the gradient of the chemical potential) for mechanical processes we integrate over the volume of the system:

$$d\phi = \int dV \sum_k J'_k dX'_k \leq 0 \ . \qquad (8)$$

## 7. Understanding entropy in a cell

Several ideas have been proposed trying to understand the production entropy inside a cell, even arguing to go against the tendency of second law, which is invalidate for a living system, because the principle that entropy can only increase or remain constant applies only to a closed system which is adiabatically isolated, meaning no heat can enter or leave. Whenever a system can exchange either heat or matter with its environment, an entropy decrease of that system is entirely compatible with the second law [1].

## 7.1 Prigogine Model

Prigogine developed a model demonstrating clearly how a non-equilibrium system can become unstable and make a transition to an oscillatory state. To understand its theoretical simplicity in the trimolecular model, we shall understand first the reaction scheme [7]:

$$A \overset{k_1}{\rightarrow} X$$

$$B + X \overset{k_2}{\rightarrow} Y + D$$

$$2X + Y \overset{k_3}{\rightarrow} 3X$$

$$X \overset{k_4}{\rightarrow} E$$

With a net reaction: $A + B \ \rightarrow \ D + E$

Summarizing, there are two kinds of reactions, the ones liberating free energy and the ones that call for free energy, which is required to perform mechanical and chemical work. In order that two reactions can be coupled it is necessary the existence of a common intermediate .

Assumptions taken: concentrations of the reactants A and B are maintained and the products D and E are removed as soon as they are formed [7]. Furthermore we assume that all the reverse reactions are so slow they can be neglected, leading to the following rate equations for the species X and Y:

$$\frac{d[X]}{dt} = k_1[A] - k_2[B][X] + k_3[X]^2[Y] - k_4[X] \equiv Z_1$$

$$\frac{d[Y]}{dt} = k_2[B][X] - k_3[X]^2[Y] \equiv Z_2$$

## 7.2 A protocell

To be able to quantify the entropy production in a living cell we need to consider simple models, which allow us to understand these reactions in the simplest possible way.

In a cell, a huge number of chemical is synthetized by mutual catalyzation leading to replication of molecules allowing cell grow until the cell is large enough to divide into two. A protocell is a model of a cell created in order to investigate how chemical compositions are synthetized by mutual catalyzation leading to replication of molecules that allow the grow of the cell until its division [10].

As an example please see below the schematic drawing (figure 3) of a protocell with a "self-producing" system described as an approach for an artificial cell [9]:

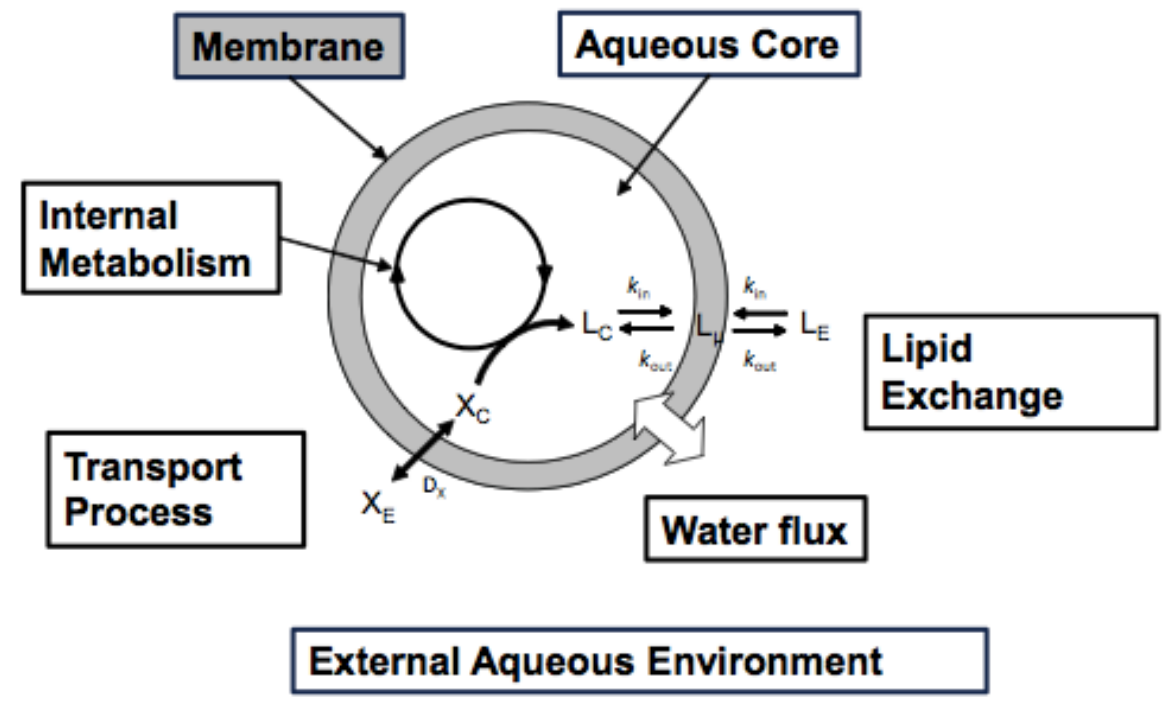

Figure 3: A protocell capable to synthesize membrane constituents. Source: [9]

The term protocell is widely used as referred to a self-organized, endogenously ordered, spherical collection of lipids, which has been proposed as the key step to understand the evolution of life (please see figure 4).

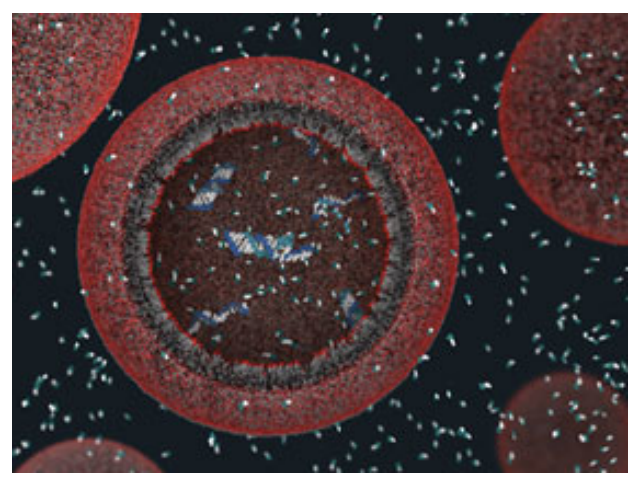

Figure 4: A three-dimensional view of a model protocell with DNA strands, shown as the ancestor of life. Source: [www.nsf.gov]

## 8. The Entropy Production of a cell

Until now only general concepts have been explained. The next step is to put together all these concepts and apply them to the idea presented in [2], to be able to explain the entropy production in a cellular growth.

In the paper three approaches have been followed, form the simplest model where the entropy production results from the intracellular reaction consisting only of the synthesis of the enzyme and membrane precursors from the nutrient.

### 8.1 Two component model

In this simple case the protocell consists only of nutrient, membrane precursor and enzyme (figure 5).

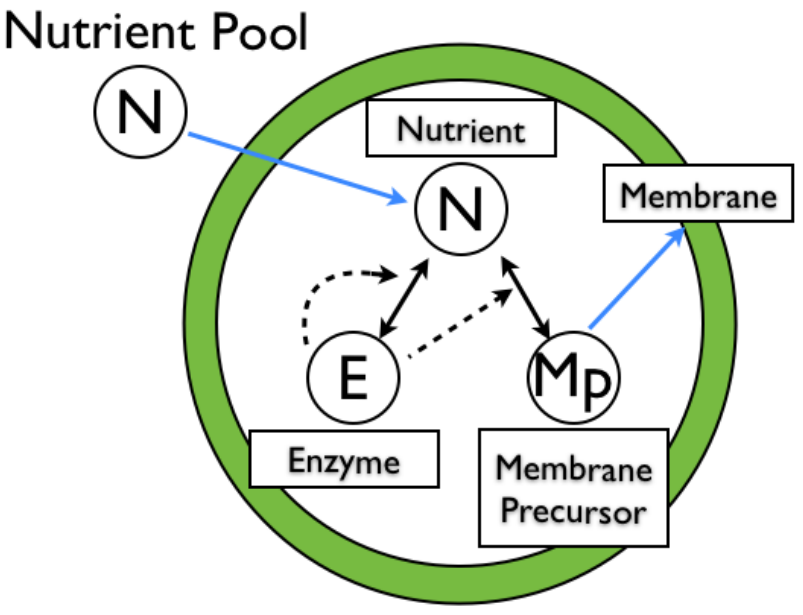

Figure 5: Simple protocell model. Source: [2]

The first step is to understand where the equations for the model come from. There are two variables: x as the concentration of the enzyme and y as the concentration of the membrane precursor.

Let's apply the rate equations from the Prigogine Model for the species X and Y, in this case, enzyme and membrane precursor:

$$\frac{d[X]}{dt} = k_1[A] - k_2[B][X] + k_3[X]^2[Y] - k_4[X]$$

$$\frac{d[Y]}{dt} = k_2[B][X] - k_3[X]^2[Y]$$

Taking the differential equations from Himeoka and Kaneko text:

$$\frac{dx}{dt} = k_x x(kX - x) - x\lambda$$

$$\frac{dy}{dt} = k_y x(lX - y) - \phi y - y\lambda$$

There is a clear resemblance between both; therefore the Himeoka and Kaneko approach is a Prigogine model with specific parameters (described in [2]).

Himeoka and Kaneko have presented the non-linear thermodynamics approach (please refer to formula 8) in the following form:

$$S = \sigma \int_0^T V_0 e^{\lambda t} dt = \frac{\sigma}{\lambda} V_0 \qquad (9)$$

with $V_0$ the initial cell volume and T is doubling time of the protocell volume.

Defining $\eta \equiv \sigma/\lambda$ as the entropy production per unit cell-volume growth, the next step is to study the dependence of η on the nutrient condition and the growth rate λ. Generally, if η is smaller, the thermodynamic efficiency for a cell growth is higher. For larger η, more energetic loss occurs in the reaction process.

In this first approach only the entropy production by the chemical reaction is considered.

The results show that the entropy production rate per unit growth η is minimal at a finite nutrient uptake rate. To explain this, the authors tell us "the extracellular concentrations of the nutrient and of the membrane precursor are far from equilibrium in the presence of catalysis". Hence with a higher nutrient uptake the entropy production rate increases but with further increases the result is instead a decrease on the nutrient uptake rate due to the cellular growth. Again, this result is in strong contrast with the thermal engine, where the entropy production is minimal at a quasi-static limit.

If the enzyme abundance is fixed to be independent of the nutrient uptake, the speed of approaching equilibrium is not altered by the nutrient condition; therefore, the entropy production just increases monotonically because of the cell volume growth.

To summarize, in a stationary state, the cell has a constant Gibbs free energy and entropy whereas in the environment the entropy increases and the free energy decreases. When considering flow the cell shows an increase in entropy and decrease in free energy, whereas both are constant in the environment.

The entropy production by the flow of chemicals from the outside of the cell will be considered in the next section.

## 8.2 Additional entropy production by material flow

In this second approach the material flow is considered as contributor for the entropy production. The authors [2] have defined s as the nutrient concentration and by including the temporal evolution of s we have the following differential equations:

$$\frac{ds}{dt} = -k_x x(ks - x) - k_y x(ls - y) - s\lambda + D(s_{ext} - s)$$

$$\frac{dx}{dt} = k_x x(ks - x) - x\lambda$$

$$\frac{dy}{dt} = k_y x(ls - y) - \phi y - y\lambda$$

The rate at which the nutrient is taken from the extracellular environment with a concentration $s_{ext}$ is denominated D in the first equation.

The results show again that under non-equilibrium chemical flow the minimal $\eta$ is achieved for a finite nutrient uptake. As the entropy production is primarily derived from chemical reactions, the conclusion of the preceding section doesn´t change.

## 8.3 Multi-component model

In this last approach it is considered to have a system with a large number of chemical species, which is the model closest to the reality, since there is a large number of chemical reactions present in a cell.

The authors [2] have defined $X_1$, $x_1$, $x_N$ and $x_i$ as the concentrations of the external nutrient, the nutrient, the membrane precursor and the enzymes in order to have the following equations:

$$\frac{dx_1}{dt} = \sum_{j=1}^{N} \sum_{k=2}^{N-1} \left( C(1,j;k) k_{1j} x_j - C(j,1;k) k_{j1} x_1 \right) x_k + (X_1 - x_1) - x_1 \lambda$$

$$\frac{dx_i}{dt} = \sum_{j=1}^{N} \sum_{k=2}^{N-1} \left( C(i,j;k) k_{ij} x_j - C(j,i;k) k_{ji} x_i \right) x_k - x_i \lambda \quad (1 < i < N-1)$$

$$\frac{dx_N}{dt} = \sum_{j=1}^{N} \sum_{k=2}^{N-1} \left( C(N,j;k) k_{Nj} x_j - C(j,N;k) k_{jN} x_N \right) x_k - \phi x_N - x_N \lambda$$

$$\lambda = x_N$$

being C the reaction tensor.

It is also important to highlight that in this model the reactions were considered as reversible.

By numerical simulations they have provided as a result the existence of an optimal point of $\eta$ for a random generated network of N=100. Conclusion, the entropy production is minimized at a non-zero nutrient concentration.

### 8.3.1 The Kullback-Leibler (KL) divergency of the model

The Kullback-Leibler (KL) divergence is a fundamental equation of information theory and statistics that quantifies the proximity of two probability distributions. It is considered that in the entire universe entropy is information and in the case of living systems the same approach can be used [11].

This divergence is also known as information divergence and relative entropy, and measures the information one random variable contains about a related random variable. In the model in [2] the steady state distribution from the equilibrium Boltzmann distribution as a function of the external nutrient concentration is calculated.

The result shows a reduction of $\eta$ and it is has been suggested that it is related to the equilibration process of abundant enzymes synthetized as a result of the high nutrient uptake discussed in the previous sections.

## 9. Summary

A cell grows through nutrient consumption described in the equations shown before in each of the model approaches.

The authors concentrate in the computation of $\eta$ (entropy production per unit cell volume growth) and found that the value is minimum at a certain nutrient uptake rate.

Nevertheless by a higher growth range the value of $\eta$ increases. This is explained by the assumptions taken at the beginning, as the main differentiation from a Carnot engine: equilibration of non equilibrium environmental conditions facilitated by the enzyme, autocatalytic processes to synthetize the enzyme and cell-volume increase resulting from membrane synthesis.

In the figure 6 a simple entropy balance for a growing cell is illustrated. This is a very good example of what it could happen inside the cell. In the model presented by Himeoka and Kaneko material loss is not included in the model but it can be assumed that energy dissipations is correlated with material dissipation. Another example very characteristic of metabolism can be found in the figure 7. This kind of model could be a good way to continue future studies.

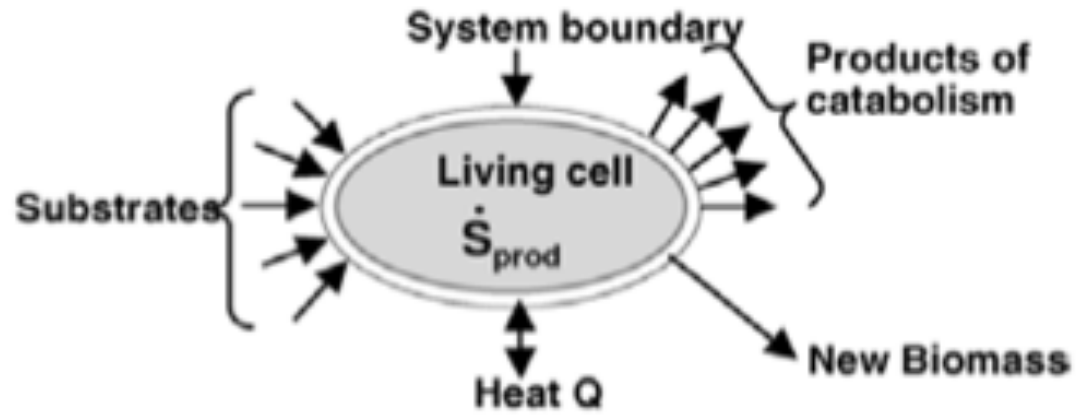

Figure 6: Entropy balance for a growing cell. Source: [4]

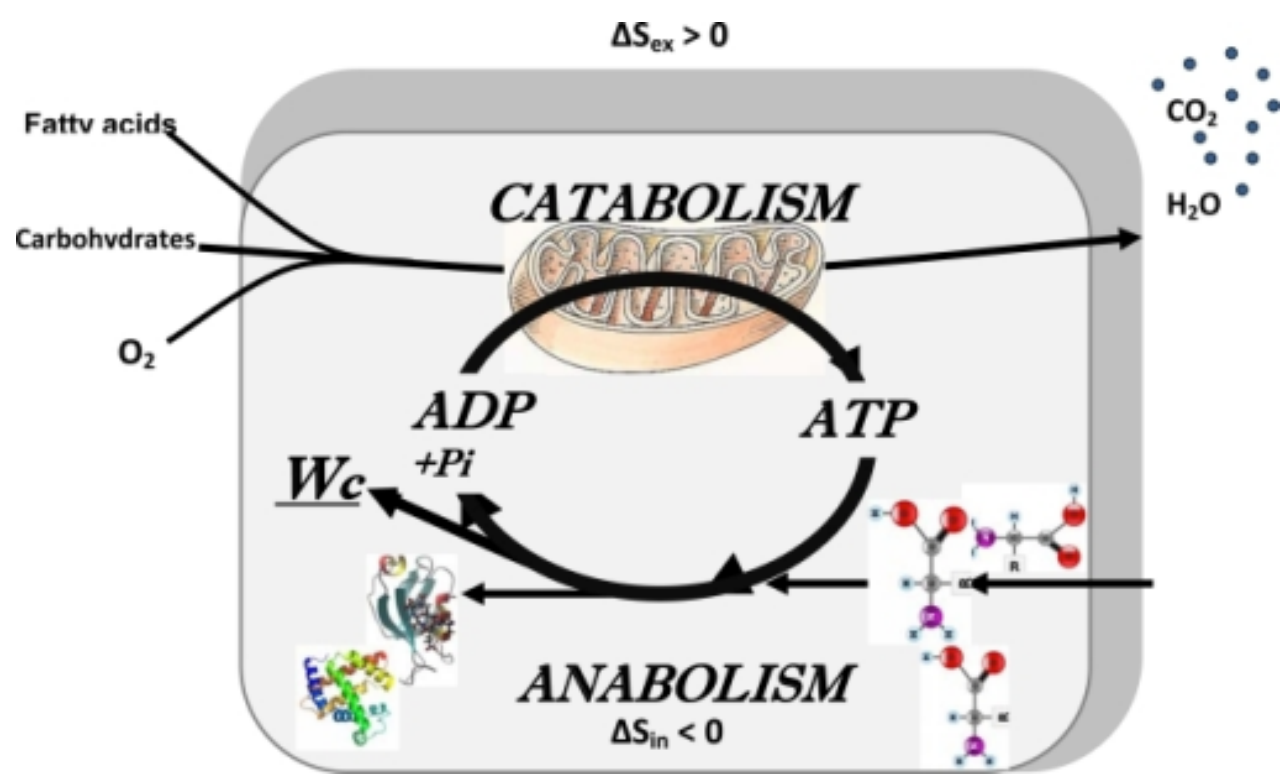

Figure 7: General scheme for cellular metabolism. Source: [www.openi.nlm.nih.gov]

Considering the correlation exposed between energy and matter, the minimal entropy production at a finite nutrient flow presented in the models could be used to find an optimal yield at a finite nutrient flow. The models discussed by the authors are simple models not taking into account the big complexity of the biochemistry inside the cell, but it is a good approximation that can be used in future models to understand the growth of the cell due to a certain nutrient uptake rate.

## 10. Discussion

In the approach followed by Himeoka and Kaneko there were a lot of assumptions taken as well as simplifications in order to have a reliable result. Again, to be able to have a complete understanding of a cell a full model taking into account of chemical reactions and energy balance have to be included.

In science we use models to understand complicated processes through simplification. In the paper here discussed I think it was really good to see the three different model approaches from the simplest one to the multi-component model.

It is very difficult to simulate living systems as models. Sometimes scientists try to simulate any process in the simplest possible way in order to understand the behaviour of cells. The question is: can we really model life? Can we really try to make a simple way to describe the most amazing feature in our universe? As illustrated through this assignment there have been several initiatives in the last century as well as more complicated elaborations in the present times.

The next step could be to realize artificial cells able to grow and divide by themselves (autopoietic cells), which would be really helpful in biology, medicine and therapy against cancer.

Entropy is a physical concept. Entropy can be explained in the vast empty space, in any engine or mechanical system, in all processes which surround us. But entropy is a characteristic not easy to measure in a living cell. If we take evolution, is it negative entropy then? Is not the disorder becoming organized? Are we then intelligent beings and therefore with more order in our lives, or is intelligence making us more entropic and chaotic?

All these questions could be answered step by step by better approximations on entropy production studies in cells.

## 11. Bibliography

- [1] Schroedinger, E; “What is life?”, Cambridge University Press, 1994.
- [2] Himeoka, Y, Kaneko, K; “Entropy production of a steady-growth cell with catalytic reactions”, Physics Rev, 2014.
- [3] Buceta, J, Koroutcheva, E, Pastor, J. M; “Temas de Biofísica”, UNED Ediciones, 2006.
- [4] Jha, P. K, Huda, S; “Entropy Generation in living systems: Cellular Scale”
- [5] Smith, E; “Thermodynamics of natural selection II: Chemical Carnot Cycles”, Journal of Theoretical Biology 252:198-212, 2008.
- [6] Davies, P. C.W, Rieper, E, Tuszynski, J.A; “Self-organization and entropy reduction in a living cell”, Biosystems, 111(1):1-10, January 2013.

- [7] Kondepudi, D, Prigogine, I; "Modern Thermodynamics: From Heat Engines to Dissipative Structures", John Wiley & Sons, 1998.
- [8] Prigogine, I; "Thermodynamics of irreversible processes", John Wiley & Sons, 1968.
- [9] Mavelli, F, Altamura, E, Cassidei, L, Stano, P; "Recent theoretical approaches to minimal artificial cells", Entropy, 16, 2488-2511, 2014.
- [10] Kondo, Y, Kaneko, K; "Growth states of catalytic reaction networks exhibiting energy metabolism", Physical Review E84, 011927, 2011.
- [11] Pérez Cruz, F; "Kullback-Leibler Divergence Estimation of Continuous Distributions", Department of Electrical Engineering Princeton University.